# On-the-fly 3D metrology of volumetric additive manufacturing

Antony Orth,[1,*] Kathleen L. Sampson,[1] Yujie, Zhang,[1] Kayley Ting,[1,2] Derek Aranguren van Egmond,[1] Kurtis Laqua,[1] Thomas Lacelle,[1] Daniel Webber,[1] Dorothy Fathi,[1] Jonathan Boisvert,[1] and Chantal Paquet[1]

[1]*National Research Council of Canada, 1200 Montreal Road, Ottawa, Ontario, Canada K1A 0R6*
[2]*University of Waterloo, Waterloo, Ontario, Canada*
*[*antony.orth@nrc-cnrc.gc.ca](mailto:antony.orth@nrc-cnrc.gc.ca)*

**Abstract:** Additive manufacturing techniques are revolutionizing product development by enabling fast turnaround from design to fabrication. However, the throughput of the rapid prototyping pipeline remains constrained by print optimization, requiring multiple iterations of fabrication and *ex-situ* metrology. Despite the need for a suitable technology, robust *in-situ* shape measurement of an entire print is not currently available with any additive manufacturing modality. Here, we address this shortcoming by demonstrating fully simultaneous 3D metrology and printing. We exploit the dramatic increase in light scattering by a photoresin during gelation for real-time 3D imaging of prints during tomographic volumetric additive manufacturing. Tomographic imaging of the light scattering density in the build volume yields quantitative, artifact-free 3D + time models of cured objects that are accurate to below 1% of the size of the print. By integrating shape measurement into the printing process, our work paves the way for next-generation rapid prototyping with real-time defect detection and correction.

## 1.  Introduction

Recent advancements in volumetric additive manufacturing (VAM) are a paradigm shift for additive manufacturing (AM). Instead of slowly building up objects layer-by-layer, VAM techniques allow for rapid 3D printing of all layers simultaneously[1–3]. Currently, the most widely used VAM approach is tomographic VAM, where the polymerizing light dose is decomposed into 2D light patterns via tomographic principles[1,3–11]. In this approach, a projector back-projects the object's (filtered) Radon transform through a rotating vial filled with photocurable resin. Once the local absorbed light dose exceeds the gelation threshold, the resin solidifies. In this way, tomographic VAM can be used to print complex parts without the layering artifacts and support structures associated with stereolithography (SLA) and digital light processing (DLP) 3D printing. Crucially, VAM techniques avoid the time-consuming peel-recoat steps in SLA and DLP, enabling sub-minute print times.

While VAM has pushed AM speed to new heights, a major remaining challenge across all AM modalities is the need for rapid print inspection and metrology[12–14]. A typical AM workflow involves an inspection step to verify print quality to assess dimensional tolerances. This step is performed after printing due to the limitations of current geometrical metrology techniques. For example, X-ray computed tomography (CT)[15] is not practical for *in-situ* implementation due to ionizing radiation precautions. Similarly, 3D laser scanning and optical profilometry require specialized optical setups that cannot be easily integrated into the build frame, and so are generally performed *ex-situ* [5]. As a result, time-consuming metrology and associated iterative print optimization are a major bottleneck in rapid prototyping, where print speed is often no longer the rate limiting factor. A real-time *in-situ* metrology technology for AM would transform the sequential print/measure/repeat workflow into a combined print *and* measure approach.

Currently, *in-situ* AM metrology is typically limited to assessing local material properties [16–19]. Recently, there have been a few reports of AM systems that measure print geometry between layer depositions via optical 3D scanning. These systems, however, either sample only a subset of layers [20,21] or assume that the print closely matches a simplistic design file *a priori* [22]. Moreover, these techniques are not truly parallel with printing, as they require the printing process to be paused in order to make a measurement. At present there is no technology that can reliably assess the geometry of the entire object *in-situ* during or after printing. Although high resolution imaging of the printed part can be performed *ex-situ*, this can be extremely time-consuming. For example, the gold standard metrology technique for AM, micro x-ray CT, typically requires on the order of an hour for data capture alone [15]. This is in addition to transport time between AM and x-ray CT facilities, and significant operational and capital expenses. Optical 3D scanners are more accessible but cannot map occluded geometries that often occur in complex prints, and only work with optically diffuse surfaces. Surface profilometry offers high resolution imaging of local surface topography, though only over approximately flat regions with small deviations from the mean.

In this work, we introduce an optical imaging technique for real-time 3D mapping of curing in AM. We take advantage of the sharp increase in light scattering by a resin as it undergoes gelation to distinguish between cured and uncured regions of the print volume [23,24]. By recording the intensity of light scattering within the build volume in a darkfield configuration, we obtain high contrast projection images of the print. Cured regions are highly scattering and appear bright, whereas the uncured monomer remains dark. Here, we apply this general concept to tomographic VAM, where the rotating build volume provides a means for tomographic dose projection *and* tomographic imaging simultaneously. We call our imaging technique "Optical Scattering Tomography" (OST) in analogy with closely related "Optical Projection Tomography" (OPT) [25–27]. In OST, we directly image side-scattered light from the print volume while it rotates (Fig. 1), building up a sinogram of each layer of the object, enabling subsequent tomographic reconstruction. To our knowledge, this is the first demonstration of a photopolymerization AM system that enables the entire print to be faithfully and quantitively mapped live during the printing process. We note that although we describe a tomographic system in this paper, the general approach of scattered-light imaging by a resin during gelation may also be applied to continuous and discrete layer-wise AM techniques such as continuous liquid interface production (CLIP) [28], inhibition patterning [29], Xolography [30], two photon photopolymerization [31], SLA and DLP.

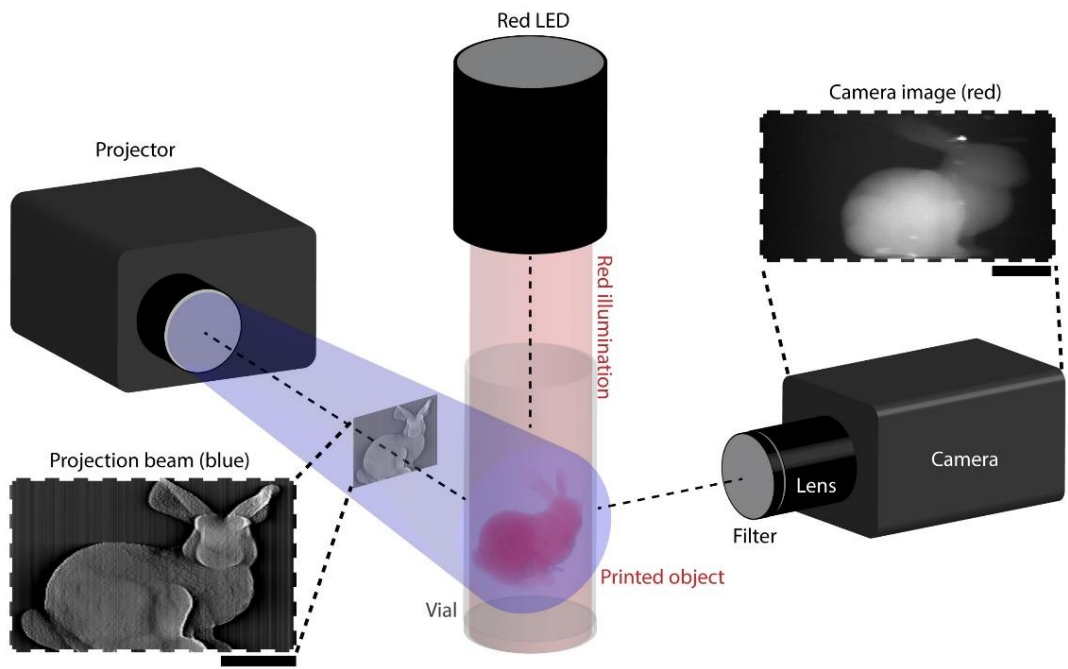

**Fig. 1**. Experimental setup for OST imaging in tomographic VAM. The projector projects patterns of blue light onto the cylindrical vial containing photocurable resin. This blue light drives the polymerization process. The projector beam inset shows an example projection frame during printing. A collimated overhead red LED illuminates the contents of the vial. The red light is outside the absorption band of the photoinitiator in the resin and therefore does not interfere with the writing process. A camera is oriented orthogonal to both the vial and the illumination LED to observe the contents of the vial. A bandpass filter centered at the wavelength of the illumination LED is placed on the camera lens to filter out all light except light from the red LED. The inset shows an example raw frame from the camera showing the emergence of a Stanford Bunny. Scalebars are 5 mm.

Beyond the need for real-time geometry monitoring across all AM techniques, VAM stands to benefit in particular. Unlike SLA and DLP, where the applied light dose outside of the boundary of the object is effectively zero, volumetric dose projection used in VAM is unable to create regions of zero dose. As a result, the projection light must be turned off immediately once the object has solidified into its desired shape. If light projection continues past this point, regions outside the desired object will start to solidify as the local accumulated light dose rises to the gelation threshold, resulting in poor printing fidelity [2]. To contend with this exposure time sensitivity in VAM, the optimal exposure time must be known. Though this practical issue is not often discussed in the VAM literature, it prevents repeatable and accurate printing results.

Previous reports have attempted to address this challenge by reconstructing properties of the print tomographically. One paper describes how a gelation time map was estimated from transmission images of the print during rotation in tomographic VAM [3]. However, reconstruction was based on a qualitative thresholding of the image and was performed off-line for subsequent optimization. The authors show improvement of the optimized print, but do not show reconstruction of the printed object or the accuracy of the technique. Another recent report describes tomographic reconstruction of the refractive index field in tomographic VAM via color-Schlieren tomography [6]. Again, this approach does not quantitatively measure the geometry of the gelled print, instead it must be estimated from the refractive index change. This differs from our work where we measure light scattering density, which we connect quantitatively to the gelation threshold. Moreover, OST is simpler to implement than color-Schlieren imaging: OST requires only a monochrome light source as opposed to broadband illumination and a Fourier-plane hue filter required by the color-Schlieren technique. Finally, whereas our approach is agnostic to the orientation of the print boundary, the demonstrated color-Schlieren technique can only resolve boundaries oriented along the direction of the 1-D color-Schlieren filter. This can potentially introduce artifacts if the print geometry primarily contains features orthogonal to the Schlieren filter.

In this paper we first outline the mathematical framework for our implementation of OST in our printer geometry. We then demonstrate 4D (3D + time) imaging of complex tomographic prints and compare OST imaging results with the reference print geometry. We find excellent agreement between experimental OST-derived 3D meshes and reference geometries, with dimensional accuracies and root-mean-square errors (RMSE) typically equal to or less than the OST voxel size, or about 1% or less of the print size. This performance indicates that our tomographic print system achieves state-of-the-art VAM print fidelity. This is remarkable given that our tomographic VAM system does not employ index matching fluid around the print volume. Instead, projected patterns *and* recorded images are corrected computationally to reverse the strong refractive artifacts introduced by the curved air/glass interface of the print vial. Moreover, OST imaging of VAM processes stands apart from other metrological methods currently available to other 3D printing processes in that it monitors objects being built from the bottom-up (molecular precursors to object). This opens the door to incorporating feedback controls for *in-situ* dose corrections/compensation, real-time quality control and standardization that are essential for the successful commercialization of VAM.

## 2. Working Principle

The print projection system used in this work was outlined in a previous report [4]. Briefly, a digital projector projects light patterns through a rotating cylindrical glass vial containing a photocurable resin. The projected patterns are constructed such that a 3D light dose distribution corresponding to the desired print emerges in the resin after an integer number of rotations. Regions of high dose (ie. inside the object region) solidify, while the dose in the remaining volume is below the gelation threshold and therefore these regions remain liquid. The key feature of our tomographic printing system is that it does not contain a refractive index matching bath around the print vial, making it simpler to build and more user-friendly. As a result, however, light rays refract significantly at the air/vial interface. In [4] we showed that this can be corrected for via

Radon-space resampling derived from a detailed ray-tracing analysis. The same challenge presents itself in the current work, where we wish to image a volume inside the vial. Light rays originating in the vial undergo strong refraction as they exit the vial. Thus, light recorded by the camera is not composed of rays travelling parallel to the optical axis, as is required in traditional tomography. Instead, the direction of the light ray varies across the field-of-view and the direction of these rays changes at the air/vial interface. In order to correct for this effect and achieve tomographic imaging reconstruction, we need to apply a resampling step to convert the as-imaged data into a standard Radon transform. This Radon transform is then inverted using a standard Fourier back-projection (FBP) method to obtain the desired 3D reconstruction of the print.

In the following subsection, we will summarize the resampling approach used to compute the projection patterns needed to print the object [4]. We will then outline the imaging geometry and the analogous image resampling procedure that allows for visualization of the object being printed in the vial.

*2.1 Projection Resampling*

In this section, we describe the resampling step in terms of finding the relationship between the coordinates of a virtual parallel beam projector and the physical projector. The virtual projector represents the case where there is no refraction at the air-vial interface. If this were the case, the desired dose in the resin could be achieved by projecting the (filtered) Radon transform (or sinogram) of the object as the vial rotates. However, rays emitted by the physical projector are subject to refraction. By calculating the location of a general light ray on both projectors, we can derive a mapping between the two coordinate systems [4]:

$$(x_p, \theta) \leftrightarrow (x_v(x_p), \theta_v(x_p, \theta)) \tag{1}$$

Here, $x_p$ is the horizontal coordinate of the physical projector and $\theta$ is the physical vial rotation angle; $x_v$ is horizontal coordinate on a virtual projector, and $\theta_v$ is the rotation angle of the virtual projector. The horizontal axis is approximately magnified, while the angular coordinate transform is more complex.

$$\theta_v = \sin^{-1}\left(\frac{x_p^*}{R_v}\right) - \sin^{-1}\left(\frac{n_1}{n_2}\sin\theta_i\right) + \theta \tag{2}$$

$$x_v = x_p^* \cos\theta_v - \sqrt{R_v^2 - x_p^{*2}}\sin\theta_v \approx \frac{n_1}{n_2}x_p \tag{3}$$

where $\theta_i$ is the angle of incidence of the light ray on the vial, $R_v$ is the vial radius, and $n_1$ and $n_2$ are the refractive indices outside and inside the vial, respectively. The variable $x_p^*$ takes into account the effect of the projector's throw ratio $T_r$:

$$x_p^* = x_p \left(1 - \sqrt{1 - \alpha(1 - (R_v/T_r W)^2)}\right)/\alpha \qquad (4)$$

where $\alpha = 1 + (x_p/T_r W)^2$, and $W$ is the width of the projector. With this remapping in hand, the FBP-filtered sinogram $S$ of the desired object is then mapped on to the physical projector space via a resampling step to obtain the resampled sinogram $S_r$. Though we use an FBP-filtered sinogram in our work, we note that projections calculated using iterative optimization techniques can be used as input to the resampling step instead [5,32]. The resampled sinogram $S_r$ is then projected through the rotating vial with the physical projector to create the desired dose profile. The remapping step assures that the projected $S_r$ are predistorted to exactly compensate for the in-plane refraction at the air/vial interface and the non-telecentricity of the projector.

We note that the combination of projector non-telecentricity and air/vial refraction imposes a small approximate compression (~0.96-0.97x) of the projected image along the axis of the vial, as described in Supplementary Information Section 1 and Fig. S1. This is compensated for by vertically stretching of the input geometry during slicing.

*2.2 Imaging*

In this section we outline the imaging geometry of our tomographic printing setup. We call this technique "Optical Scattering Tomography" (OST) due to its tomographic nature and the scattering contrast mechanism. As shown in Fig. 1, the print vial is illuminated from above with an approximately collimated red LED source (SugarCube Red, 624nm center wavelength). A camera (FLIR USB3 Grasshopper) outfitted with a lens (Edmund Optics 25mm/F1.8 #86572) and a 624nm bandpass filter images side scattered light from the contents of the vial. The optical axes of the camera, the projector, and red-light source are all mutually orthogonal. As the vial rotates during printing, the camera records integrated scattering density projections through ray trajectories in the vial as shown in Fig. 2a. This can be understood by tracing an arbitrary backwards-propagating ray from the camera back through the vial. Every voxel in the vial intercepted by this ray is a potential scattering site that can contribute to the signal at the camera. The collection of side scattering images of the vial over a full rotation comprises a full set of tomographic data suitable for 3D reconstruction of the scattering density. However, the imaging distortions caused by refraction at the air/vial interface and the non-telecentricity of the imaging system violate the parallel beam assumption needed for FBP reconstruction. As with pattern projection, the raw scattering images need to be resampled to remove this distortion before FBP reconstruction.

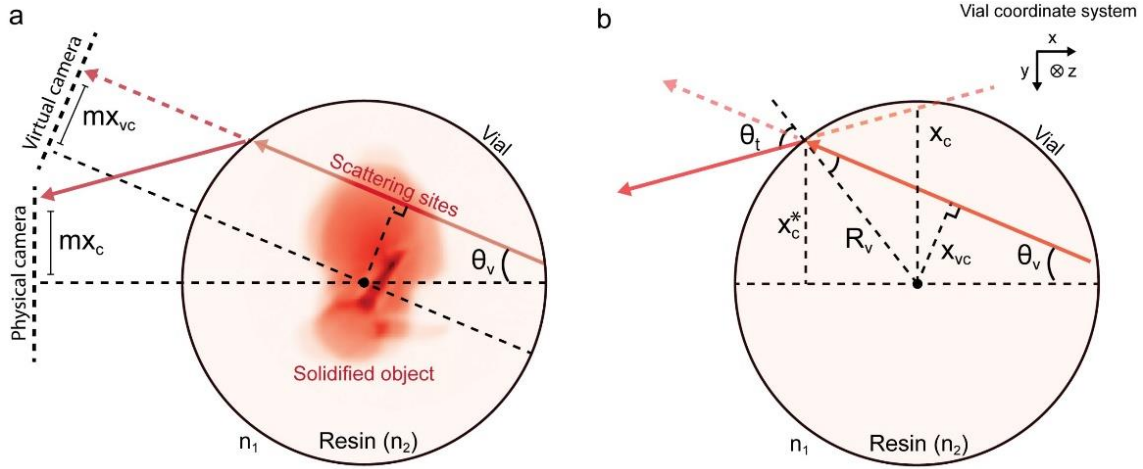

**Fig. 2**. Ray tracing geometry. a) Ray tracing diagram showing the trajectory of a red light scattered by the object back to a (real) physical camera and a virtual camera. The red ray in the vial indicates possible scattering sites that contribute to the signal recorded at a particular pixel on the physical and virtual cameras at the positions indicated by the solid and dashed red rays outside the vial, respectively. Here the scalar factor $m$ refers to the magnification factor of the camera + lens system. The solidified object acts as a source for scattering events though its entire volume. b) Same as in (a) but with the object removed and geometric variables defined.

### 2.3 Image resampling

A ray tracing diagram for an arbitrary ray impinging on a camera pixel is shown in Fig. 2b. This diagram is analogous to that in the projection case from [4], as required by time-reversal symmetry. For imaging, physical and virtual camera coordinates $x_c$ and $x_{vc}$ replace the projector coordinates $x_p$ and $x_v$, and the angle of incidence $\theta_i$ is now the angle of transmission $\theta_t$. Furthermore, the non-telecentricity on the imaging side is more conveniently described using the distance from the camera to the vial centre $D$ instead of the projector throw ratio. The remapping for imaging can be found by a substituting the physical and virtual camera coordinates $x_c$ and $x_{vc}$ for the projector coordinates $x_p$ and $x_v$ in Equations 2-4:

$$\theta_v = \sin^{-1}\left(\frac{x_c^*}{R_v}\right) - \sin^{-1}\left(\frac{n_1}{n_2}\sin\theta_t\right) + \theta \tag{5}$$

$$x_{vc} = x_c^* \cos\theta_v - \sqrt{R_v^2 - x_c^{*2}}\sin\theta_v \approx \frac{n_1}{n_2}x_c \tag{6}$$

where,

$$x_c^* = x_c \left(1 - \sqrt{1 - \alpha_i \left(1 - \left(\frac{R_v}{x_c}\tan(x_c/D)\right)^2\right)}\right)/\alpha_i \tag{7}$$

$$\alpha_i = 1 + (\tan(x_c/D))^2 \tag{8}$$

$$\theta_t = \sin^{-1}\left(\frac{x_c^*}{R_v}\right) + \tan^{-1}\left(\frac{x_c^*}{D}\right) \tag{9}$$

*2.4 Tomographic reconstruction*

With the imaging remapping in hand, the 3D scattering density in the vial can be reconstructed. This process is illustrated in Fig. 3. Figures 3a-c show images during a print of a Stanford Bunny with diurethane dimethacrylate (DUDMA) resin at three different vial rotation angles during a full vial rotation (angular sampling step is 2°). From the stack of images of a full rotation, a distorted sinogram is extracted for each z-slice. One such slice is shown in Fig. 3d, where the location of the slice is indicated by the dashed horizontal lines in Figs. 3a-c. This sinogram, which is distorted by vial refraction and non-telecentricity and sampled in $(x_c, \theta_v)$-space, is then remapped to $(x_{vc}, \theta)$-space (Radon space) to produce the undistorted sinogram shown in Fig. 3e. This sinogram is then back-projected via FBP to create a reconstructed OST slice in vial space (Fig. 3f). This reconstructed slice represents the scattering density within this 2D slice of the vial. The 3D scattering density field in the vial is obtained by repeating this process for all horizontal slices in the print vial.

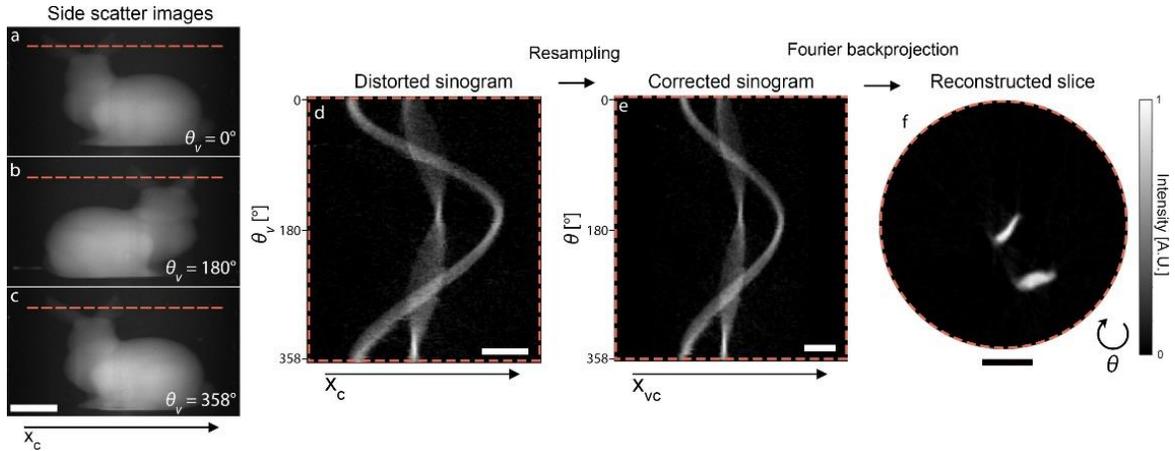

**Fig. 3**: OST reconstruction process. (a)-(c) Raw side scatter images of a Stanford Bunny print at 3 different rotation angles indicated by $\theta_v$. The dashed horizontal line indicates the plane shown in (d)-(f). This plane intersects the middle portion of the bunny ears. d) The raw sinogram from the collection of side scatter images. This sinogram is distorted due to air-vial refraction and non-telecentricity of the camera lens. e) Corrected sinogram, constructed by remapping

the distorted sinogram in (d) to virtual camera space via Equations 5-9.  f)  The reconstructed slice is calculated by Fourier backprojecting the corrected sinogram in (e).  Bright regions in this slice correspond to solidified portions of the bunny ears.  Scale bars are 5mm.

Analogous to the projection analysis in Supplementary Information Section 1, there is a small approximate expansion (~1.03x) of the imaged geometry along the vertical dimension.  This is corrected for by a vertical demagnification of the geometry after tomographic reconstruction.

## 3.  Results

A volumetric visualization of the 3D scattering field for the Stanford bunny[33] print in Fig. 3 is shown in Fig. 4a, along with an overhead sum projection image in Fig. 4b.  Although these data capture the volumetric nature of the imaging data, it is not the most informative way to visualize the print geometry.  Instead, an isosurface rendering gives far better appreciation of surface geometry (Fig. 4c).  For each resin used, we find the OST intensity value corresponding to the gelation threshold, which we call $I_p$ (see Section S2, Figs. S2-4 for details), by printing a standard cylinder.  By setting the isosurface threshold at $I_p$, we render the surface corresponding to the actual polymerized object.  If the isosurface threshold is set below this value, we may obtain an isosurface larger than the actual print, and vice versa for a smaller isosurface threshold.  Alternatively, the object is visualized by thresholding the scattering volume at $I_p$ and setting voxels over this value to 1.  The subsequent overhead sum projection of this thresholded volume is shown in Fig. 4d, which is also shown to the operator in real-time.  The print termination time is chosen by the operator based on when this live visualization of the print qualitatively matches the known reference geometry.

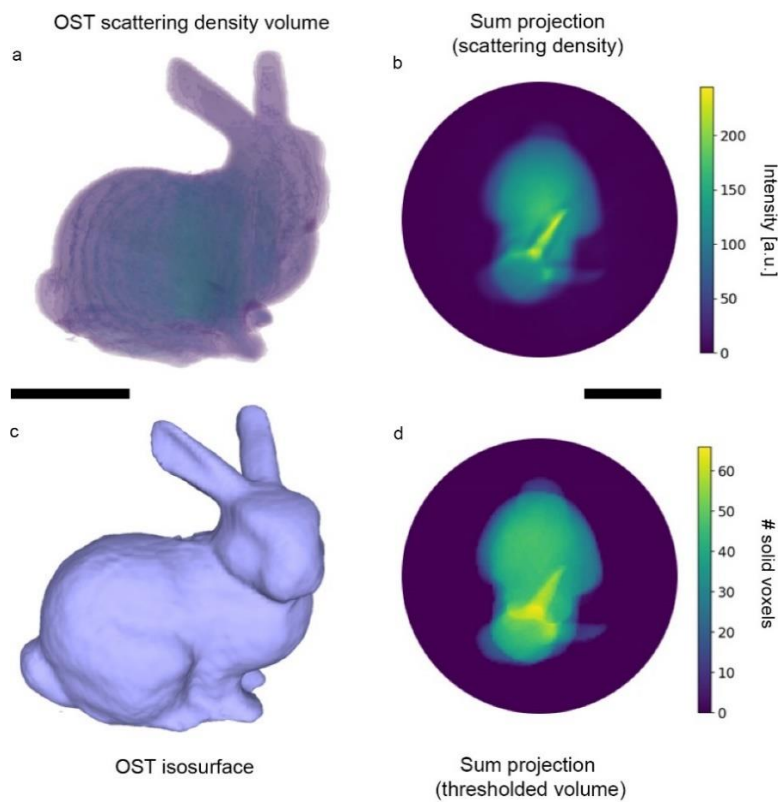

**Fig. 4**. OST print reconstruction. a) Visualization of the tomographically reconstructed OST scattering density volume for a Stanford Bunny print. b) Overhead sum-projection of the volume in (a). Features such as ears, head, body and tail are readily identifiable. c) Rendering of an OST isosurface of the Stanford Bunny print. Subtle features are much more easily visible with this surface rendering compared to a volumetric rendering. d) Overhead sum-projection of the isosurface in (c). Scalebars are 5mm.

The temporal resolution afforded by our implementation of OST uncovers interesting polymerization dynamics at play in tomographic printing. Figure 5 shows the evolution of a Benchy print during rotations 12-18 (216s-324s). The projector beam is turned off after 17 rotations (306s), with data from the 18[th] rotation yielding the final print geometry once polymerization had ceased. This series of renderings demonstrate what we found to be a common occurrence across all geometries: larger features (such as the boat hull) appear first, followed by fine features (such as the thin walls of the boat cockpit). We suspect that this is due to a combination of oxygen diffusion and optical point spread function effects, though a thorough investigation and correction is beyond the scope of this paper. To the right of each isosurface rendering in Fig. 5, we show the overhead sum projection of the OST volume for each timepoint. From this series of overhead projections, the operator can clearly observe missing features at early print times and adjust the total print time as needed. OST volumes are computed asynchronously in a parallel core from pattern projection and updated as soon as complete. Typical volume computation time is ~8-9s, corresponding to approximately half of a vial rotation. In future work, these volumes may be used to update projections on-the-fly, and the volume reconstruction time could be significantly reduced using a graphics processing unit (GPU) and/or deep neural network techniques[34].

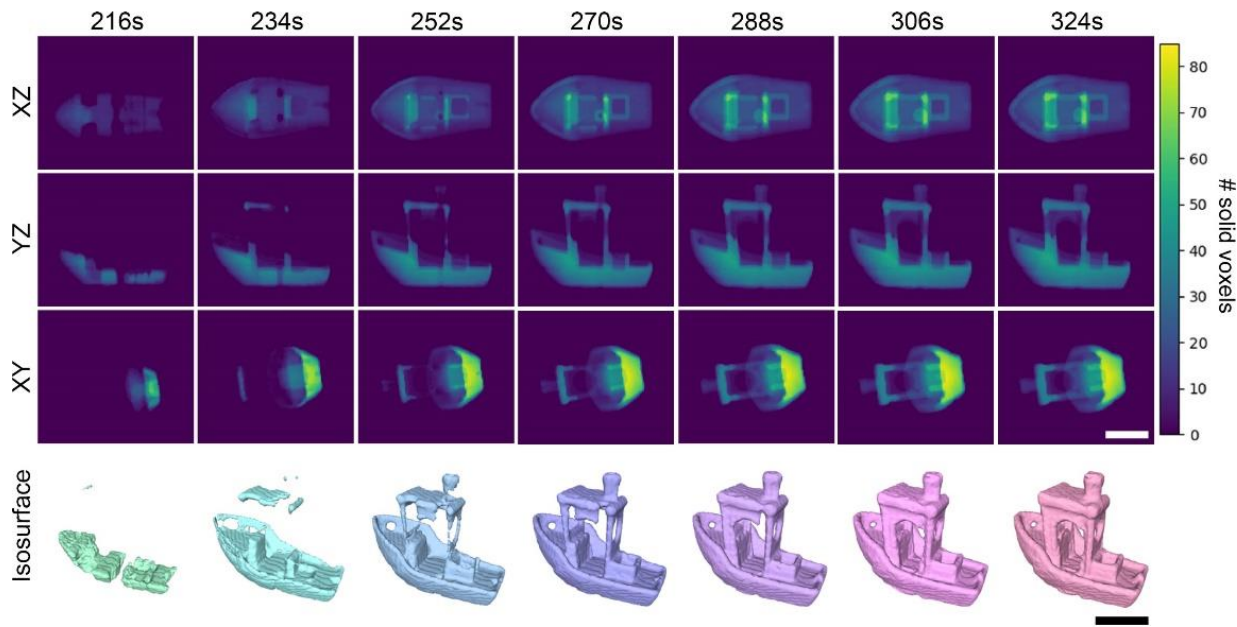

**Fig. 5.** Print dynamics are captured by OST. OST sum projections (XZ, YZ, and XY directions) and isosurface renderings of a Benchy print for rotations 12-18 (216s-324s). Large structures are the first to form and high spatial frequency structures such as the vertical supports and trim for the cockpit are the last to polymerize. The projections are displayed live to the user next to those for the reference geometry (not shown here), providing feedback about the current printed geometry. The projector is turned off after the 17th rotation (306s). Scale bars are 5mm.

In Fig. 6 we show OST isosurfaces for a variety of completed prints, along with optical photographs, and reference meshes. To provide a comparison between the reference object file and the printed object, we compute the signed distance function (SDF) between the reference meshes and OST isosurfaces in Figs. 6c,g,k and the associated histograms in Fig. S2c. The SDF is the distance between two aligned meshes at each surface patch and can be thought of as the error in the printed part geometry[35]. The full histogram of SDF values for all prints in Fig. 6 are shown in Fig. S5. The RMS (root-mean-square) of the SDF over the entire isosurface is a convenient quantification of the overall error of the print. For the Stanford Bunny print (Figs. 6a-d, DUDMA resin), we obtain an RMS error (RMSE) of 0.176mm, which is only slightly larger than the voxel size of the reconstructed mesh (0.155mm), and 1.45% of the maximum dimension of the print. Gyroid (Figs. 6e-h, BPAGDA resin) and Benchy (Figs. 6i-l, BPAGDA resin) prints yielded smaller RMSE values of 0.115mm (0.81%) and 0.158mm (1.28%), respectively, indicating excellent agreement between the OST isosurfaces from real-time reconstructed volumes and the reference design. This accuracy exceeds that recently reported in a high fidelity, index-matched system in terms of absolute RMSE values and matches those results in terms of accuracy as a percentage of the maximum print dimension (see Table S1) [5]. We attribute the high print fidelity to a combination of improved projection calculation (See Methods and Fig. S6, similar to but distinct from[32]), detailed projection optics modelling including non-telecentricity correction[4], and improved process control afforded by OST imaging. The RMSE values for these prints are only ~1.5-3x the projected pixel size in air (~0.065mm), underscoring the high fidelity of the printing process relative to the scale of print resolution. We obtain similar SDF RMSE values for SDFs computed between the OST data and x-ray CT scans of the printed part (see Fig. S7). However, artifacts due to sample mounting constraints in the x-ray CT system (Fig. S8) make SDF computation between x-ray CT and OST unreliable, especially near the bottom of the print (Fig. S7e).

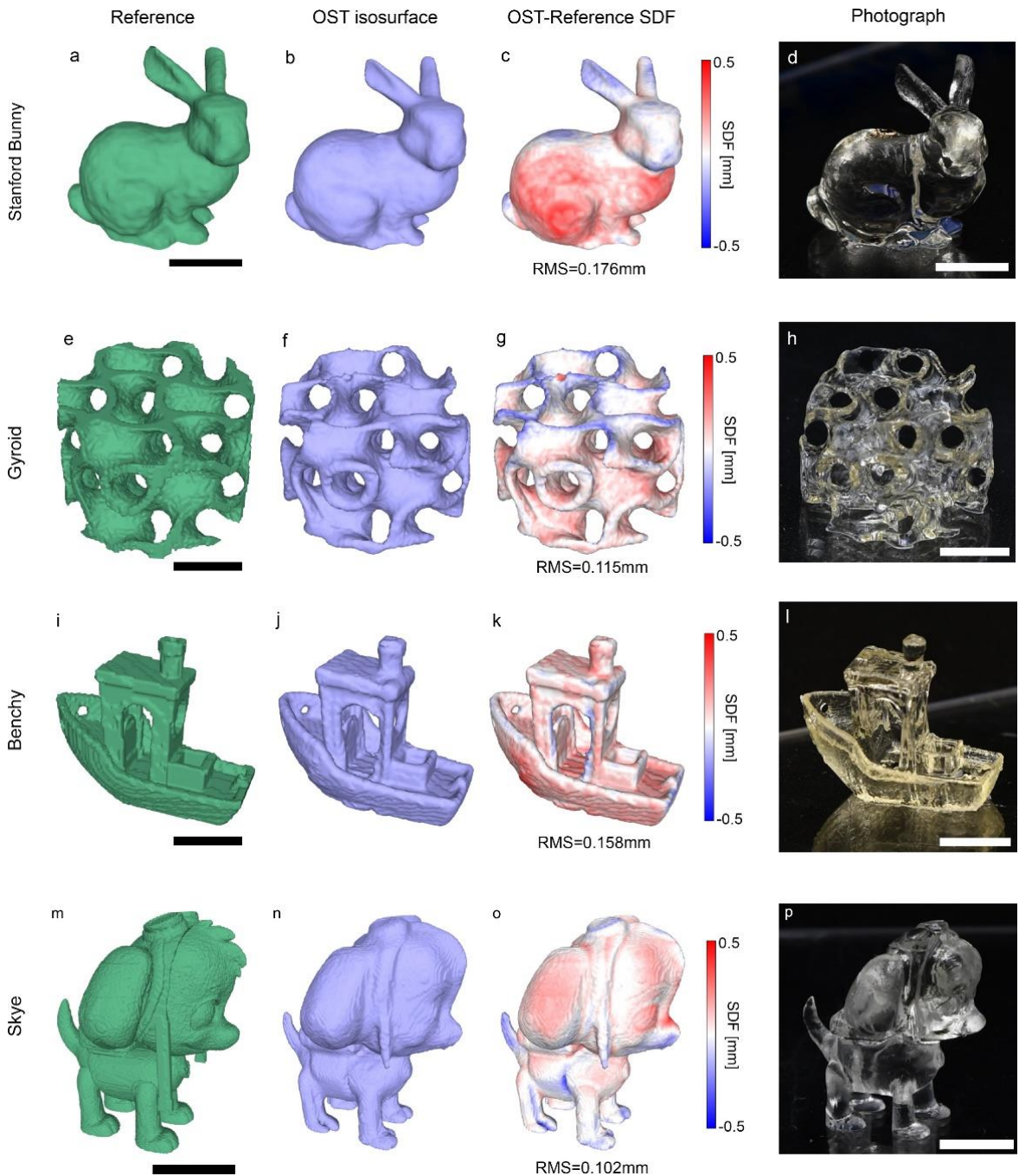

**Fig. 6.** Visualizations of various complex printed objects. a) Rendering of the reference Stanford Bunny mesh used for printing. b) OST isosurface rendering of printed Stanford Bunny. c) Signed distance function (SDF) between the OST isosurface in (b) and the reference mesh in (a). RMS of the SDF is indicated below. d) Optical photograph of the Stanford Bunny print. e)-h) As in (a)-(d), for a Gyroid geometry. The gyroid wall thickness is 1mm. i)-l) As in (a)-(d) for a Benchy print. m)-p) As in (a)-(d) for a Skye™ print. Scale bars are 5mm.

In this work we prioritized reconstruction speed and SNR over resolution to provide timely feedback to the user. To this end we downsample the image stream by 4x4 pixels, leading to the 0.155mm voxel size in reconstructed OST volume space. At this level of downsampling, we were able to resolve neighbouring lattice struts as close as 3 voxels = 0.465mm apart (Fig. S9).

Higher resolution OST volumes yielding more accurate isosurfaces can be reconstructed by relaxing the downsampling applied to the raw images. Due to the increased reconstruction time required, the volume reconstruction at this sampling density is done offline, though it could likely be done in real time with the aid of a GPU. Figures 6(m)-(p) show an OST isosurface for a Skye™ print[36] at 2x2 camera downsampling (OST voxel size 0.0775mm), with an SDF RMSE value of 0.102mm (0.70%).

## 4. Discussion

In this work we have introduced a new imaging modality called OST that enables live 3D imaging of the tomographic VAM process. OST uses the scattering arising from the micro-scale refractive index mismatch between liquid monomer and solid polymer as an optical contrast mechanism. The 3D reconstruction of the scattering density inside the print vial is enabled by tomographic sampling and a resampling process similar to that used in the projection step for a non-index-matched tomographic printing system [4]. We demonstrated that the scattering density that corresponds to gelation can be reliably calibrated, resulting in print reconstruction accurate to within 1% of the size of the print by using a physically motivated isosurface threshold. We found that even significantly convoluted objects such as the gyroid in Figs. 6e-h pose no difficulty for tomographic reconstruction via OST, making it a powerful and versatile metrology technique. Our *in-situ* imaging approach is distinct from a previously reported Schlieren-based imaging method in that OST measures scattering, whereas the Schlieren system measures ray deflection due to refractive index variation [6]. We also note that because OST relies on scattering contrast as opposed to ray deflection, more strongly scattering resins are also compatible with OST. Finally, while Schlieren-based approaches are appropriate for accurately index-matched systems, they cannot be used in non-index-matched systems such as ours due to the strength of the refraction at the air/vial interface.

OST imaging further improves ease of use for tomographic printers by providing crucial feedback of the progress of the print to the operator. We found that the OST imaging system significantly sped up our ability to obtain successful print parameters across a variety of print geometries and resins (BPAGDA and DUDMA). This was because OST eliminates the need for time consuming *ex-situ* metrology such as x-ray CT (3D), laser scanning (3D) or profilometry (2D + height). These metrology options have further specific drawbacks: in our experience, x-ray CT was time consuming and suffered from artifacts from the mounting substrate, giving an unreliable 3D model near the bottom of the printed object (Figs. S7 and S8). Complex objects, such as the gyroid in Figs. 6e-h are not even amenable to *ex-situ* laser scanning because not all the surfaces are optically accessible to the laser scanner due to occlusion. Object painting or coating for

3D laser scanning, which is necessary to make the print surface diffuse, also poses challenges for complex geometries such as non-uniform deposition. Profilometry, meanwhile, is only suitable for simple 2D surface measurements. OST alleviates all these issues, yielding remarkably defect-free models of prints throughout the entire print procedure.

The real-time nature of OST print imaging also allows the user to disentangle the source of print failures by pinpointing *when* and *where* it occurred during printing, something which is not currently possible in VAM, SLA or DLP. OST also eliminates trial and error associated with finding new print times when using new resins. Furthermore, we found that OST enabled the operator to compensate for printing speed changes due to temperature variation and when reusing resin (though all prints in this work were done on fresh, room temperature resin). In addition to being a vital tool for end user printer operation, OST may also find application in studying photopolymerization kinetics [2,23,37]. This will be key to optimizing resins and achieving industry-standard print fidelity in VAM.

The key feature of the metrology approach introduced in this paper is darkfield imaging of scattered light during the gelation process. While we have demonstrated the utility of this optical arrangement in tomographic VAM, it may also be used in other AM techniques. For example, darkfield scattering from each layer of a DLP or SLA print could be imaged from below the build volume, resulting in high contrast layer-by-layer images of the actual cured print.

Our introduction of OST sets tomographic VAM apart from other 3D printing techniques by enabling the user to visualize and quantitatively measure the print as it forms. We believe that convenient spatial-temporal metrology like that provided by OST will be critical to improving the fidelity of tomographic VAM while making it vastly more approachable for end-users.

## 5. Methods

**Projection calculation**

The input STL file is first sliced and rasterized using the Trimesh python package [38], yielding a stack of binary images $I_i$ representing the geometry of the object where values of 0 and 1 represent empty and solid regions, respectively (Fig. S6a). The image stack is then converted to floating point data type and pixels with value 0 are set to a background level $B = 0.5$, yielding a new stack of images $I_{0,i}$ (Fig. S6d). We call these images the initial target dose. The purpose of the background level is to compensate for negative intensities introduced during Fourier filtering.

The Radon transform is then calculated for each slice $i$ of the initial target dose, with angular sampling of 1°, yielding a stack of sinograms. These sinograms are then ramp filtered in Fourier space, followed by setting all values in real space below the background level to $B$. For each slice $i$, the filtered, non-negative sinogram $S_{0,i}$ is then back projected to find the corresponding simulated dose $D_{0,i}$. We subsequently calculate a dose normalization coefficient for each slice: $N_{0,i} = D_{0,i}/I_{0,i}$ and set the next iteration of the target dose: $I_{1,i} = I_{0,i}/N_{0,i}$. A new set of filtered, non-negative sinograms $S_{1,i}$ and their associated dose $D_{1,i}$ are then calculated for the updated target dose stack $I_{1,i}$. The normalization step is repeated, yielding $N_{1,i} = D_{1,i}/I_{0,i}$ and $I_{2,i} = I_{0,i}/N_{1,i}$. This iterative process can be repeated indefinitely to create a high contrast set of non-negative sinograms that closely approximate the original target dose $I_{0,i}$. Because we wanted to minimize projection computation time, we only performed one such iteration. See Fig. S6 for an example of the projection calculation process. For all prints in this work, we use ramp filtered, non-negative sinograms of the image stack $I_{1,i} = I_{0,i}/N_{0,i}$ (a slice of one such image stack is shown in Fig. S6g). Figure S6 shows the effect of the background level on the simulated dose as well as the effect of the single target dose modification iteration. Both steps help to improve dose uniformity as can be seen by the simulation dose maps (Figs. S6b,e,h) and their associated histograms (Figs. S6c,f,i).

Before the resampling step, we also correct for the light penetration depth in the vial. For computational simplicity, we take a different approach from that reported in [1]. Instead of computing the exponential radon transform at the outset, we instead simulate the dose $D_d = D_d(x, y)$ delivered when attempting to print a uniform disk with radius equal to the radius of the addressable write volume. The simulated light dose is lowest at the center of the vial and highest at the edge of the disk, as expected. We then use this 2D simulated dose profile as a normalization map for a slice of the image stack $I_{1,i}$ described above. That is, we calculate an absorption corrected image stack $I_{1,i}^c = I_{1,i}/D_d$. This process corrects for systematic underexposure at the center of the write volume due to blue light absorption in the vial. Such a correction is not required on the imaging side because the resin does not absorb appreciably in the red.

**Printing**

Resin-filled vials were centered on a rotation stage (Physik Instrumente M-060.PD) with a custom designed vial holder. The position of the vial in the field of view of the projector was measured by sweeping a vertical line horizontally across the projector field and capturing the photoinitiator fluorescence with the camera. This alignment step is only completed once and does not need to be updated unless the system comes out of alignment.

Projections are calculated and resampled according to the above "Projection calculation" method. Once this step is complete, the projection patterns are projected through the vial at a rate of 32 fps. To avoid issues with unsteady projector frame rate, the python script polls the computer for the current time at each iteration of a loop. The appropriate frame to display is then found based on the elapsed time from the start of projection.

Simultaneous with the start of pattern projection, rotation stage motion is initiated at a speed of 20°/s, and proceeds for 20 rotations. Prints are terminated manually by the operator by placing a beam block in front of the projector prior to 20 rotations if needed.

**Imaging**

Raw OST image frames (i.e. Figs.3a-c) are captured via a FLIR USB3 Grasshopper camera (GS3-U3-23S6M-C) in the arrangement shown in Fig. 1. The camera is fitted with a c-mount lens (Edmund Optics 25mm/F1.8 #86572), with the aperture set to F/8. A spectral bandpass filter (624nm, FWHM 10nm) was placed in front of the lens that passes only the red LED light.

The camera was set to operate in "Mode 5", which provides 4x4 pixel binning from the image stream (note, binning is not applied at the physical pixel level). Images were captured at 10 frames per second, corresponding to an angular sampling step of 2°. This yields a spatial sampling step of $2\pi R_v \frac{n_1}{n_2} \times 2°/360° = 283 \mu m$ at the edge of the printable area in the vial ($R_v = 12.4$mm as measured by digital calipers).

**Post-processing**

Finished prints were removed from the vial with a metal spatula and placed immediately in a dish filled with isopropyl alcohol (IPA). The print was left to soak in IPA for 10-20 mins and then removed and left to dry at room temperature. Prints were subsequently post-cured using 405nm light for 120 mins at 60°C in a Formlabs Form Cure.

**Profilometry**

Cylinder samples were profiled with a Cyber Technologies CT100 white light optical profilometer (Figs. S2a, S3a, S4a,c). Pixel size was set to $50 \mu m$.

**Gyroid design**

The cylindrical gyroid lattice in Figs. 6e-h and S7 - a type of triply periodic minimal surface (TPMS) - was modelled algorithmically in nTopology software (New York, USA). Here, the gyroid geometry is represented as the solution to an implicit level set equation, enabling rapid design iterations and rendering of the topology. A TPMS sheet thickness of 1.0mm was chosen for the curved surfaces, while the repeating unit cell size was set to 7mm. The continuous TPMS field was intersected with a solid cylinder of 14mm diameter and 10mm height to yield the final cylindrical gyroid. Finally, the model was converted into a surface mesh with 0.5mm beam elements and exported in STL file format for printing.

**Resin**

*DUDMA*

This resin was prepared by mixing diurethane dimethacrylate (DUDMA) with poly(ethylene glycol) diacrylate (Mn=700 g/mol, PEGDA 700) in a 8:2 weight ratio. The photoinitator system includes camphorquinone (CQ) and ethyl 4-dimethylaminobenzoate (EDAB). CQ is used at a concentration of 7.8mM, and the same weight of EDAB is used. All chemicals above are purchased from Sigma Aldrich and used as received. The prepared resins were stored at 4°C until needed. Before printing, resin was decanted into open top glass vials used for printing (nominal diameter 25mm, measured diameter 24.8mm, Kimble) and allowed to warm to room temperature. If any residual bubbles remained, the resin was left to sit until the bubbles had disappeared.

The refractive index of the liquid resin was measured to be 1.49 at 460nm and 1.48 at 624nm using a Schmidt-Haensch ATR-BR refractometer.

The room temperature viscosity of the DUDMA resin was measured to be 1,100 cp using a Brookfield DV-III Ultra Programmable Rheometer.

*BPAGDA*

The resin was prepared similarly to that reported previously in literature (all materials were purchased from Sigma Aldrich (Milwaukee, United States) and used as received) [1,4]. Two acrylate crosslinkers were used as the precursor materials: bisphenol A glycerolate (1 glycerol/phenol) diacrylate [BPAGDA] and poly(ethylene glycol) diacrylate Mn 250 g/mol [PEGDA250] in a ratio of 3:1. To this BPAGDA/PEGDA250 mixture, the two component photoinitiator system, camphorquinone [CQ] and ethyl 4-dimethylaminobenzoate [EDAB], was added in a 1:1 weight ratio and CQ at a concentration of 7.8 mM in the resin. The concentration of the photoinitiators was adjusted to this value such that the penetration depth of the resin was in-line with the radius of the vial. The resin was mixed using a planetary mixer at 2000 rpm for 15 min followed by 2200 rpm for 30 sec, then separated into 20 mL scintillation vials (filled to ~15 mL). The resin was kept in the fridge for storage. The decanting process was the same as for the DUDMA resin above.

The refractive index of the liquid resin was measured to be 1.55 at 460nm and 1.53 at 624nm using a Schmidt-Haensch ATR-P refractometer.

The room temperature viscosity of the BPAGDA resin was measured to be 9,000 cp using a Brookfield DV-III Ultra Programmable Rheometer.

**Signed distance function calculation**

Signed distance functions (SDFs) were calculated using MeshLab[35]. Reference and OST isosurface STL files were imported in the MeshLab and aligned using point-based estimation. The SDF between the OST isosurface and the reference was then calculated using the sampling filter menu. Each SDF was then displayed in a blue-white-red colormap on the OST isosurface and exported in PLY format to extract histogram data.

**X-ray CT**

X-ray imaging of printed samples was performed on a GE Medical Systems RS-9, using a 10 minute scan protocol. Angle increment was 0.9 degrees, pixel size in the reconstruction was $93 \mu m$, frame averaging was set to 3 with exposure time 100ms. Voltage peak was set to 80kV with current $450 \mu A$.

## 6. References

[1]  B.E. Kelly, I. Bhattacharya, H. Heidari, M. Shusteff, C.M. Spadaccini, H.K. Taylor, Volumetric additive manufacturing via tomographic reconstruction, Science. 363 (2019) 1075–1079.

[2]  M. Shusteff, A.E.M. Browar, B.E. Kelly, J. Henriksson, T.H. Weisgraber, R.M. Panas, N.X. Fang, C.M. Spadaccini, One-step volumetric additive manufacturing of complex polymer structures, Science Advances. 3 (2017) eaao5496. https://doi.org/10.1126/sciadv.aao5496.

[3]  D. Loterie, P. Delrot, C. Moser, High-resolution tomographic volumetric additive manufacturing, Nature Communications. 11 (2020) 852. https://doi.org/10.1038/s41467-020-14630-4.

[4]  A. Orth, K.L. Sampson, K. Ting, K. Ting, J. Boisvert, C. Paquet, Correcting ray distortion in tomographic additive manufacturing, Opt. Express, OE. 29 (2021) 11037–11054. https://doi.org/10.1364/OE.419795.

[5]  I. Bhattacharya, J. Toombs, H. Taylor, High fidelity volumetric additive manufacturing, Additive Manufacturing. 47 (2021) 102299. https://doi.org/10.1016/j.addma.2021.102299.

[6]  C. Chung Li, J. Toombs, H. Taylor, Tomographic color Schlieren refractive index mapping for computed axial lithography, in: Symposium on Computational Fabrication, Association for Computing Machinery, New York, NY, USA, 2020: pp. 1–7. https://doi.org/10.1145/3424630.3425421.

[7]  J. Madrid-Wolff, A. Boniface, D. Loterie, P. Delrot, C. Moser, Light-based Volumetric Additive Manufacturing in Scattering Resins, ArXiv:2105.14952 [Physics]. (2021). http://arxiv.org/abs/2105.14952 (accessed September 17, 2021).

[8]  C.C. Cook, E.J. Fong, J.J. Schwartz, D.H. Porcincula, A.C. Kaczmarek, J.S. Oakdale, B.D. Moran, K.M. Champley, C.M. Rackson, A. Muralidharan, R.R. McLeod, M. Shusteff, Highly Tunable Thiol-Ene Photoresins for Volumetric Additive Manufacturing, Advanced Materials. 32 (2020) 2003376. https://doi.org/10.1002/adma.202003376.

[9]  P.N. Bernal, P. Delrot, D. Loterie, Y. Li, J. Malda, C. Moser, R. Levato, Volumetric bioprinting of complex living-tissue constructs within seconds, Advanced Materials. 31 (2019) 1904209.

[10] M. Kollep, G. Konstantinou, J. Madrid-Wolff, A. Boniface, P.V.W. Sasikumar, G. Blugan, P. Delrot, D. Loterie, C. Moser, Tomographic Volumetric Additive Manufacturing of Silicon Oxycarbide Ceramics, ArXiv:2109.12680 [Physics]. (2021). http://arxiv.org/abs/2109.12680 (accessed October 25, 2021).

[11] J. Toombs, M. Luitz, C. Cook, S. Jenne, C.C. Li, B. Rapp, F. Kotz-Helmer, H. Taylor, Volumetric Additive Manufacturing of Silica Glass with Microscale Computed Axial Lithography, ArXiv:2110.01651 [Physics]. (2021). http://arxiv.org/abs/2110.01651 (accessed October 25, 2021).

[12] H.D. Vora, S. Sanyal, A comprehensive review: metrology in additive manufacturing and 3D printing technology, Prog Addit Manuf. 5 (2020) 319–353. https://doi.org/10.1007/s40964-020-00142-6.

[13] S.K. Everton, M. Hirsch, P. Stravroulakis, R.K. Leach, A.T. Clare, Review of in-situ process monitoring and in-situ metrology for metal additive manufacturing, Materials & Design. 95 (2016) 431–445. https://doi.org/10.1016/j.matdes.2016.01.099.

[14] L. Koester, H. Taheri, L.J. Bond, D. Barnard, J. Gray, Additive manufacturing metrology: State of the art and needs assessment, AIP Conference Proceedings. 1706 (2016) 130001. https://doi.org/10.1063/1.4940604.

[15] A. du Plessis, I. Yadroitsev, I. Yadroitsava, S.G. Le Roux, X-Ray Microcomputed Tomography in Additive Manufacturing: A Review of the Current Technology and Applications, 3D Printing and Additive Manufacturing. 5 (2018) 227–247. https://doi.org/10.1089/3dp.2018.0060.

[16] E. Binega, L. Yang, H. Sohn, J.C.P. Cheng, Online Geometry Monitoring During Directed Energy Deposition Additive Manufacturing Using Laser Line Scanning, Precision Engineering. (2021). https://doi.org/10.1016/j.precisioneng.2021.09.005.

[17] P. Charalampous, I. Kostavelis, C. Kopsacheilis, D. Tzovaras, Vision-based real-time monitoring of extrusion additive manufacturing processes for automatic manufacturing error detection, Int J Adv Manuf Technol. 115 (2021) 3859–3872. https://doi.org/10.1007/s00170-021-07419-2.

[18] A. Dijkshoorn, P. Neuvel, S. Stramigioli, G. Krijnen, In-Situ Monitoring of Layer-Wise Fabrication by Electrical Resistance Measurements in 3D Printing, in: 2020 IEEE SENSORS, 2020: pp. 1–4. https://doi.org/10.1109/SENSORS47125.2020.9278632.

[19] M. Ogunsanya, J. Isichei, S.K. Parupelli, S. Desai, Y. Cai, In-situ Droplet Monitoring of Inkjet 3D Printing Process using Image Analysis and Machine Learning Models, Procedia Manufacturing. 53 (2021) 427–434. https://doi.org/10.1016/j.promfg.2021.06.045.

[20] S. Nuchitprasitchai, M. Roggemann, J.M. Pearce, Factors effecting real-time optical monitoring of fused filament 3D printing, Prog Addit Manuf. 2 (2017) 133–149. https://doi.org/10.1007/s40964-017-0027-x.

[21] W. Lin, H. Shen, J. Fu, S. Wu, Online quality monitoring in material extrusion additive manufacturing processes based on laser scanning technology, Precision Engineering. 60 (2019) 76–84. https://doi.org/10.1016/j.precisioneng.2019.06.004.

[22] A. Malik, H. Lhachemi, J. Ploennigs, A. Ba, R. Shorten, An Application of 3D Model Reconstruction and Augmented Reality for Real-Time Monitoring of Additive Manufacturing, Procedia CIRP. 81 (2019) 346–351. https://doi.org/10.1016/j.procir.2019.03.060.

[23] T. Norisuye, M. Shibayama, S. Nomura, Time-resolved light scattering study on the gelation process of poly(N-isopropyl acrylamide), Polymer. 39 (1998) 2769–2775. https://doi.org/10.1016/S0032-3861(97)00596-X.

[24] T. Norisuye, M. Shibayama, R. Tamaki, Y. Chujo, Time-Resolved Dynamic Light Scattering Studies on Gelation Process of Organic−Inorganic Polymer Hybrids, Macromolecules. 32 (1999) 1528–1533. https://doi.org/10.1021/ma981306h.

[25] J. Sharpe, U. Ahlgren, P. Perry, B. Hill, A. Ross, J. Hecksher-Sørensen, R. Baldock, D. Davidson, Optical Projection Tomography as a Tool for 3D Microscopy and Gene Expression Studies, Science. 296 (2002) 541–545. https://doi.org/10.1126/science.1068206.

[26] J.R. Walls, J.G. Sled, J. Sharpe, R.M. Henkelman, Correction of artefacts in optical projection tomography, Phys. Med. Biol. 50 (2005) 4645–4665. https://doi.org/10.1088/0031-9155/50/19/015.

[27] A.K. Trull, J. van der Horst, W.J. Palenstijn, L.J. van Vliet, T. van Leeuwen, J. Kalkman, Point spread function based image reconstruction in optical projection tomography, Phys. Med. Biol. 62 (2017) 7784–7797. https://doi.org/10.1088/1361-6560/aa8945.

[28] J.R. Tumbleston, D. Shirvanyants, N. Ermoshkin, R. Janusziewicz, A.R. Johnson, D. Kelly, K. Chen, R. Pinschmidt, J.P. Rolland, A. Ermoshkin, Continuous liquid interface production of 3D objects, Science. 347 (2015) 1349–1352.

[29] M.P. de Beer, H.L. van der Laan, M.A. Cole, R.J. Whelan, M.A. Burns, T.F. Scott, Rapid, continuous additive manufacturing by volumetric polymerization inhibition patterning, Science Advances. 5 (2019) eaau8723. https://doi.org/10.1126/sciadv.aau8723.

[30] M. Regehly, Y. Garmshausen, M. Reuter, N.F. König, E. Israel, D.P. Kelly, C.-Y. Chou, K. Koch, B. Asfari, S. Hecht, Xolography for linear volumetric 3D printing, Nature. 588 (2020) 620–624. https://doi.org/10.1038/s41586-020-3029-7.

[31] T. Gissibl, S. Thiele, A. Herkommer, H. Giessen, Two-photon direct laser writing of ultracompact multi-lens objectives, Nature Photonics. 10 (2016) 554–560. https://doi.org/10.1038/nphoton.2016.121.

[32] C.M. Rackson, K.M. Champley, J.T. Toombs, E.J. Fong, V. Bansal, H.K. Taylor, M. Shusteff, R.R. McLeod, Object-space optimization of tomographic reconstructions for additive manufacturing, Additive Manufacturing. 48 (2021) 102367. https://doi.org/10.1016/j.addma.2021.102367.

[33] The Stanford 3D Scanning Repository, (n.d.). http://graphics.stanford.edu/data/3Dscanrep/ (accessed December 16, 2020).

[34] A. Goy, G. Rughoobur, S. Li, K. Arthur, A.I. Akinwande, G. Barbastathis, High-resolution limited-angle phase tomography of dense layered objects using deep neural networks, PNAS. 116 (2019) 19848–19856. https://doi.org/10.1073/pnas.1821378116.

[35] P. Cignoni, M. Callieri, M. Corsini, M. Dellepiane, F. Ganovelli, G. Ranzuglia, MeshLab: an Open-Source Mesh Processing Tool, Eurographics Italian Chapter Conference. (2008) 8 pages. https://doi.org/10.2312/LOCALCHAPTEREVENTS/ITALCHAP/ITALIANCHAPCONF2008/129-136.

[36] Thingiverse.com, Skye (Paw Patrol) by Gunnar86, (n.d.). https://www.thingiverse.com/thing:2070914 (accessed October 25, 2021).

[37] B. Deore, K.L. Sampson, T. Lacelle, N. Kredentser, J. Lefebvre, L.S. Young, J. Hyland, R.E. Amaya, J. Tanha, P.R.L. Malenfant, H.W. de Haan, C. Paquet, Direct printing of functional 3D objects using polymerization-induced phase separation, Nature Communications. 12 (2021) 55. https://doi.org/10.1038/s41467-020-20256-3.

[38] M. Dawson-Haggerty, Trimesh, (2020). https://github.com/mikedh/trimesh (accessed December 15, 2020).

**Acknowledgments**

The authors thank Guy Lamouche, Maxime Rivard, Paul Finnie, Zygmunt Jakubek and Hendrick de Haan for stimulating discussions.

**Funding**

Funding provided by the National research Council of Canada.

**Contributions**

AO wrote printer control software, performed prints, profilometry, and data analysis. A.O., D.W and T.L built the printer system. K.T. wrote the slicing and projection calculation software for the printer. K.L.S. and Y.Z. designed, manufactured, and characterized resins. D.A.V.E. and K.L. designed the gyroid geometry and assisted with printing. D.F. performed x-ray CT measurements. A.O., C.P., and J. B. conceptualized the technique with assistance from D.W. C.P. and J.B. supervised the research. A.O. wrote the manuscript with input from all authors.

**Data availability**

Data underlying this paper are available upon reasonable request.

**Supplementary Information for:**

**On-the-fly 3D metrology of volumetric additive manufacturing**

*Section S1: Vertical magnification factor due to non-telecentricity of projector and imaging camera, and refraction at the vial.*

A ray-tracing diagram for an off-axis projector ray travelling at an angle $\varphi_1$ with respect to the optical axis is shown in Fig. S1. Figure S1 is shown for the midplane of the vial containing the axis of rotation of the vial. We present this analysis in the midplane of the vial for simplicity, but we note that for the small ray angles encountered in this work, this analysis is also approximately valid away from the midplane of the vial.

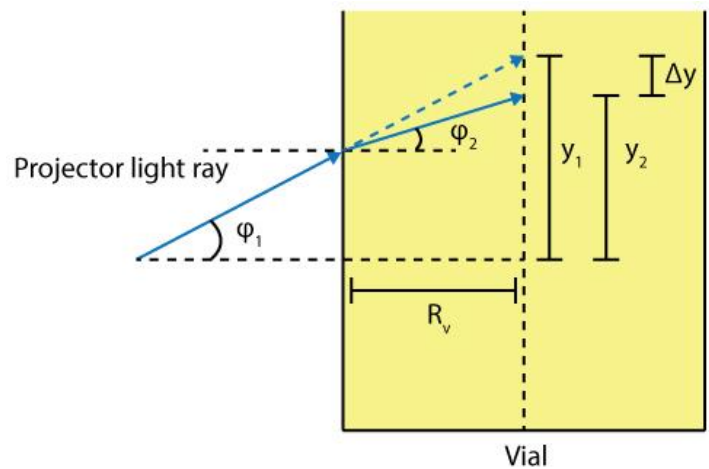

**Fig. S1**. Ray tracing diagram for a light ray impinging on the vial at a non-zero vertical angle $\varphi_1$.

If this projector light ray were to propagate without refraction, it would intersect the axis of the vial at a height $y_1$. In reality, the ray refracts at the air/vial interface and is transmitted into the vial and resin at an angle $\varphi_2$, intersecting the vial axis at a height $y_2$. We can express the height difference as

$$\Delta y = y_1 - y_2 = R_v(\tan \varphi_1 - \tan \varphi_2) \tag{S1}$$

Under the small angle approximation, we can write

$$y_1 - y_2 \approx R_v(\sin \varphi_1 - \sin \varphi_2) \tag{S2}$$

Using Snell's law and a second application of the small angle approximation, this reduces to

$$y_1 - y_2 \approx R_v \sin \varphi_1 \left(1 - 1/n_2\right) \approx R_v \tan \varphi_1 \left(1 - 1/n_2\right) \tag{S3}$$

Where $n_2$ is the refractive index of the resin. $\tan \varphi_1$ can be expressed in terms of projector parameters as $\tan \varphi_1 = \frac{y_1}{T_r W}$, where $T_r$ is the throw ratio of the projector ($T_r = 1.8$ in our system), and $W$ is the width of the projector ($W = 1024$ pixels). Using this relationship, we can rewrite Eq. S3 as:

$$y_2 \approx y_1 \left(1 - \frac{R_v(1 - 1/n_2)}{T_r W}\right) \tag{S4}$$

Finally, the vertical magnification factor $M_{v,p}$ of the image projected into the vial is:

$$M_{v,p} \equiv y_2/y_1 \approx 1 - \frac{R_v(1-1/n_2)}{T_r W} \tag{S5}$$

For our system, we have $M_{v,p} \approx 0.96$ for BPAGDA resin, and $M_{v,p} \approx 0.97$ for DUDMA resin – that is that the projected patterns are slightly smaller than intended along the vertical direction. To counteract this effect, the design geometry is stretched by a factor of $M_{v,p}$ during the initial slicing process.

This magnification effect also appears on the imaging side, where the image of the print is stretched vertically. The expression for the vertical imaging magnification $M_{v,i}$ can be obtained by considering the distance between the camera and vial $D$ instead of the projector throw ratio:

$$M_{v,i} \approx \left(1 - \frac{R_v(1-1/n_2)}{D}\right)^{-1} \tag{S6}$$

For BPAGDA and DUDMA resins, $M_{v,i} \approx 1.03$.

### Section S2: Isosurface threshold OST intensity measurement – finding $I_p$

To find the reconstructed scattering intensity $I_p$ corresponding to the gelation threshold, we print a 12mm diameter cylinder (which we call a standard) in BPAGDA and DUDMA resins. As with all prints, we use visual feedback from the real-time reconstructed OST volumes to choose when to turn off the polymerization light source. The printed cylinders are post-processed as described in Materials and Methods and subsequently imaged with a white light optical profilometer (Figs. S2a and S3a). From the optical profilometry data, the cylinder diameter is extracted using a Hough transform circle finding routine, with the resulting circle displayed in red in Figs. S2a and S3a. This circle (obtained from optical profilometry data) is then overlayed over the OST data, as shown in Figs. S2b and S3b. The gelation threshold intensity $I_p$ is set to the mean OST intensity value of all pixels intercepted by the optical profilometry-derived circles (Figs. S2c and S3c). We note that the histograms exhibit a bimodal distribution about the mean, which is likely due to a slight decenter of the circle from the measured OST data.

The validity of this approach is evidenced in the main text by the low RMSE values of the thresholded OST meshes compared to the reference mesh. To further demonstrate the accuracy of this method, we print two DUDMA validation cylinders and measure their diameters in the same way as for the standard cylinders. In both cases, we find that the difference between the optical profilometry-derived and OST-derived diameters is between 0.1-0.2mm (Fig. S4), in accordance with the RMSE values in Table S1 for more complex objects. These validation cylinders provide further evidence that our approach results in quantitatively accurate 3D models of printed objects.

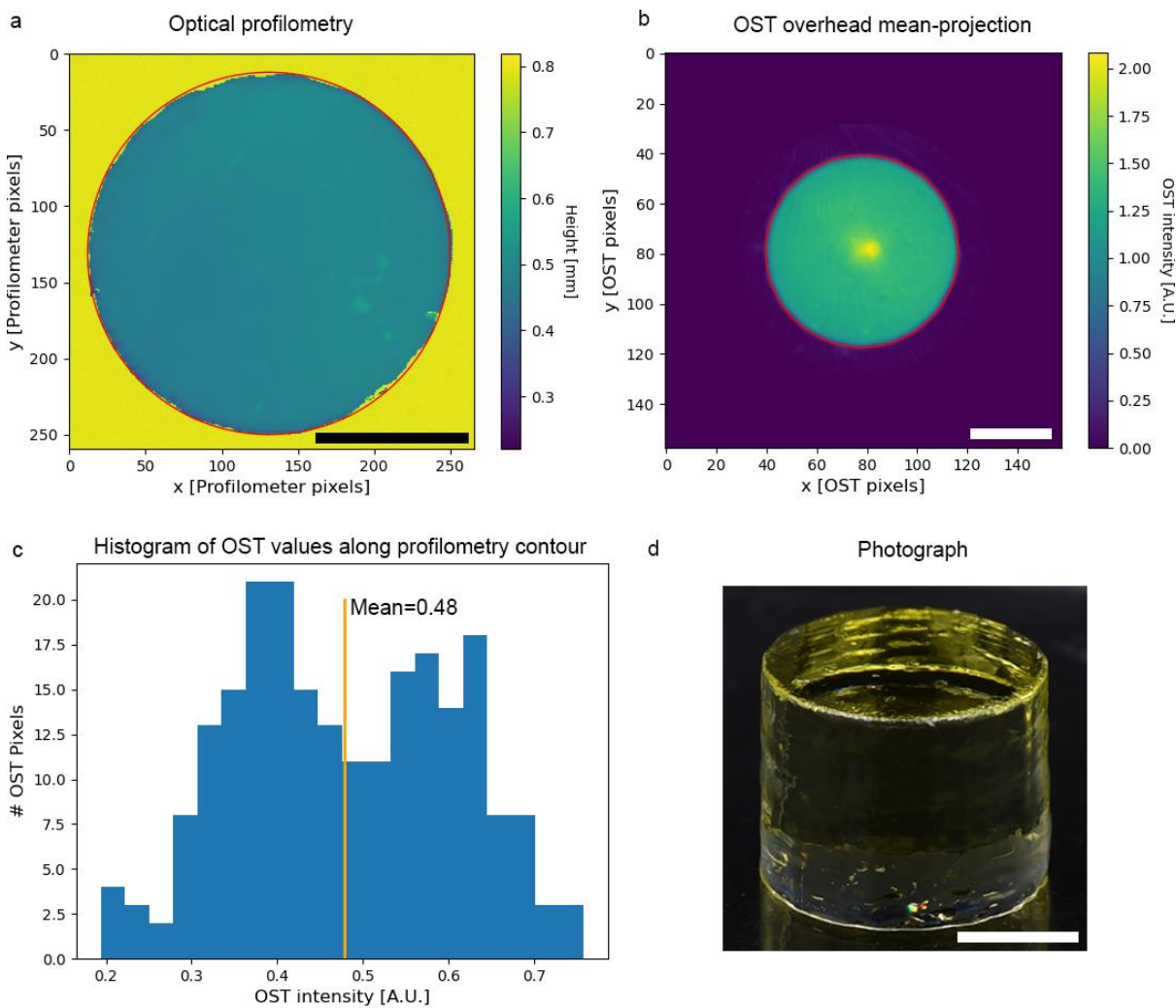

**Fig. S2**. a) Optical profilometry scan of a reference cylinder (BPAGDA resin) with design diameter 12mm. The red circle indicates the circle contour as measured by a Hough transform circle-finding routine (diameter = 11.90mm). b) OST overhead mean-projection of the printed cylinder. The red circle indicates the circle contour found in (a) via optical profilometry. c) Histogram of OST overhead mean-projection values in (b) that intersect the red circle. The gelation threshold in units of OST scattering intensity is found by calculating the mean of this distribution (OST intensity = 0.48 = $I_{p,BPAGDA}$). d) Photograph of the printed cylinder. The height cylinder's OST isosurface is 9.86mm (compared to a design value of 10mm). Scale bars are 5mm.

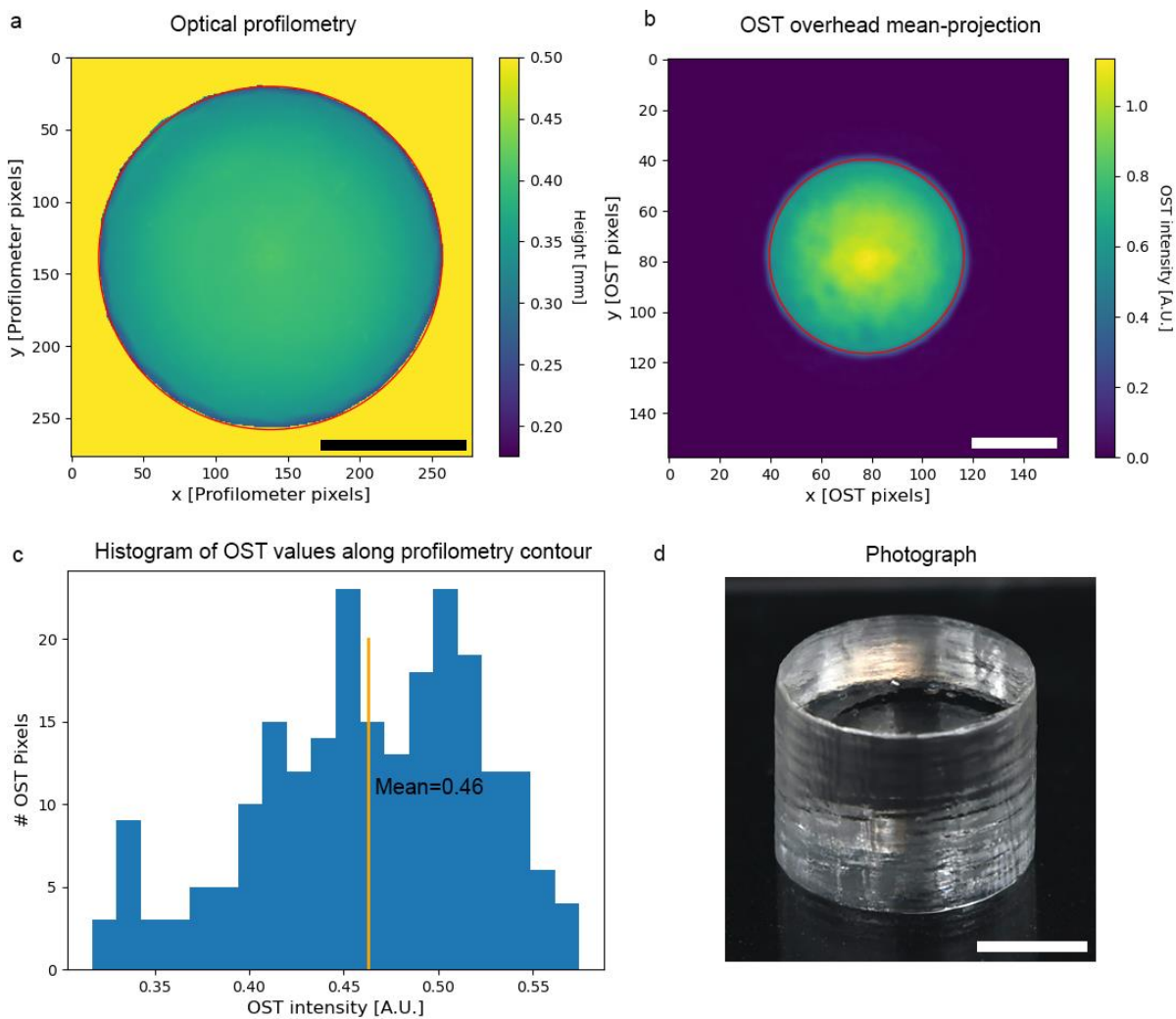

**Fig. S3**. a) Optical profilometry scan of a reference cylinder (DUDMA resin) with design diameter 12mm. The red circle indicates the circle contour as measured by a Hough transform circle-finding routine (diameter = 11.90mm). b) OST overhead mean-projection of the printed cylinder. The red circle indicates the circle contour found in (a) via optical profilometry. c) Histogram of OST overhead mean-projection values in (b) that intersect the red circle. The gelation threshold in units of OST scattering intensity is found by calculating the mean of this distribution (OST intensity = 0.46 = $I_{p,DUDMA}$). d) Photograph of the printed cylinder. Scale bars are 5mm.

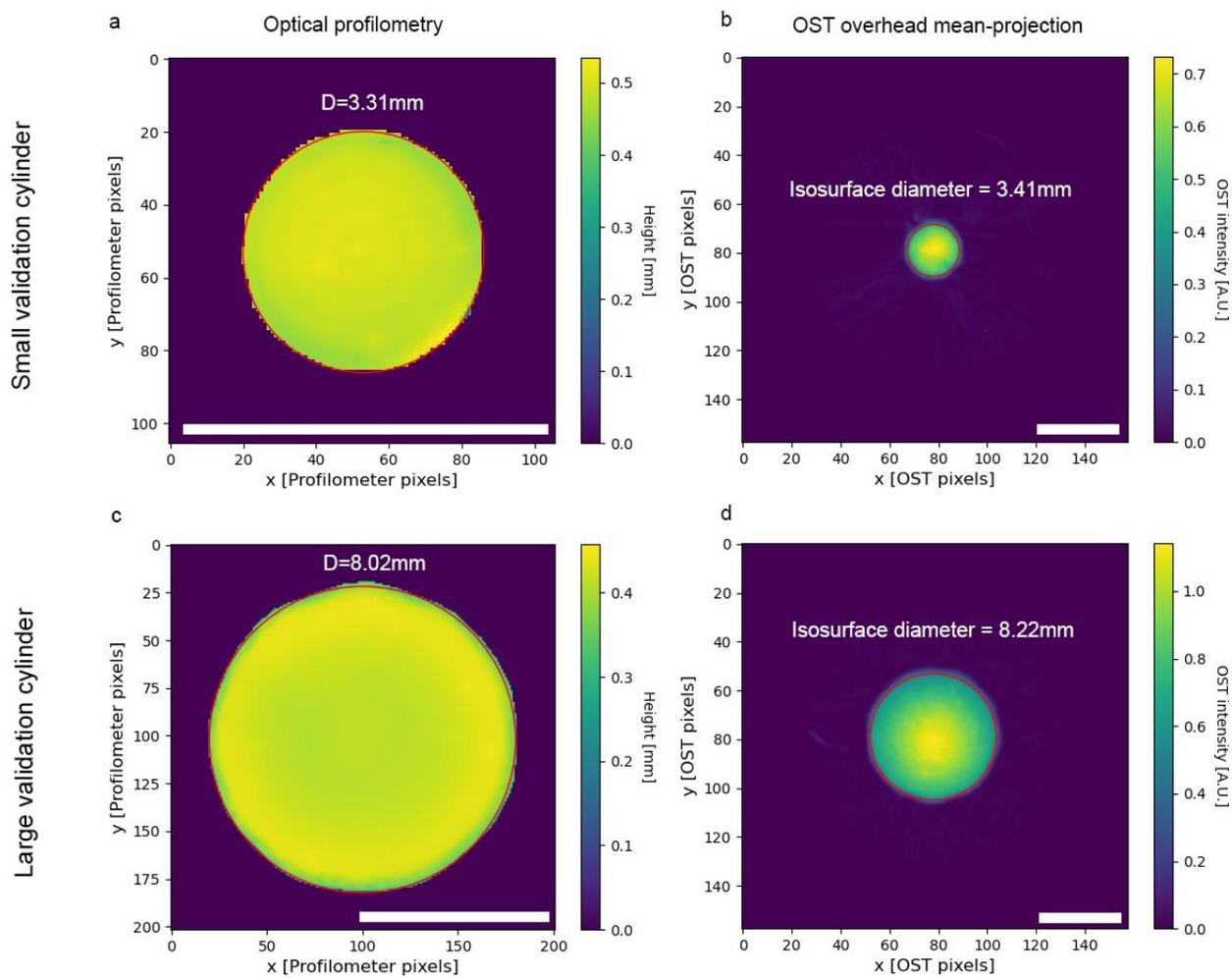

**Fig. S4**. Isosurface validation cylinders.  a)  Optical profilometry scan of a small cylinder printed in DUDMA resin. The diameter of the top of the cylinder as measured in the profilometer is D=3.31mm.  b) OST overhead projection of the top of the small validation cylinder in (a).  The diameter of the OST isosurface is 3.41mm.  The isosurface threshold value $I_{p,DUDMA}$ was taken from Fig. S3.  c) and d) As in (a)-(b) but for a larger test cylinder.  The OST isosurface diameters match those measured with profilometry to within 0.1-0.2mm, similar to the typical RMSE values for OST-to-reference values measured for complex shapes (see Table S1).  These measurements indicate that the $I_p$ values found in Figs. S2-3 are appropriate for finding the true surface geometry of the print from the OST data.

## Section S3: *Signed distance functions and print accuracy*

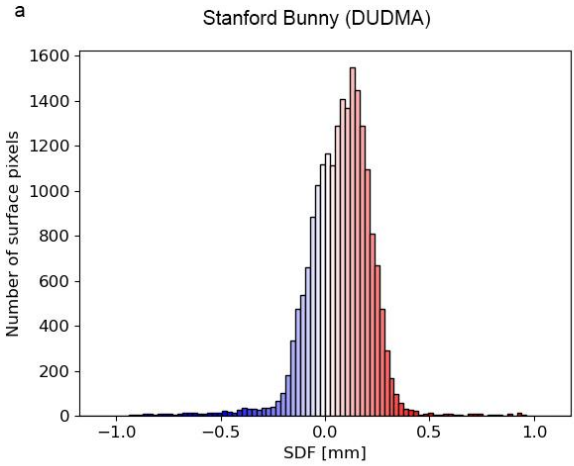
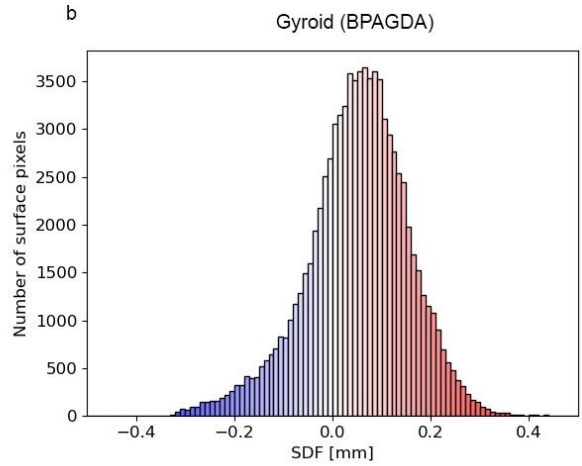
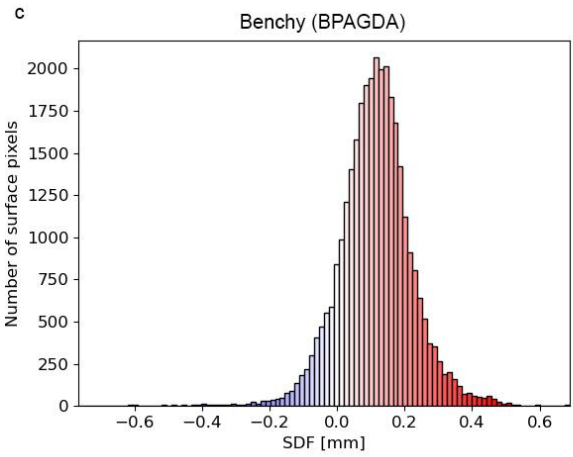
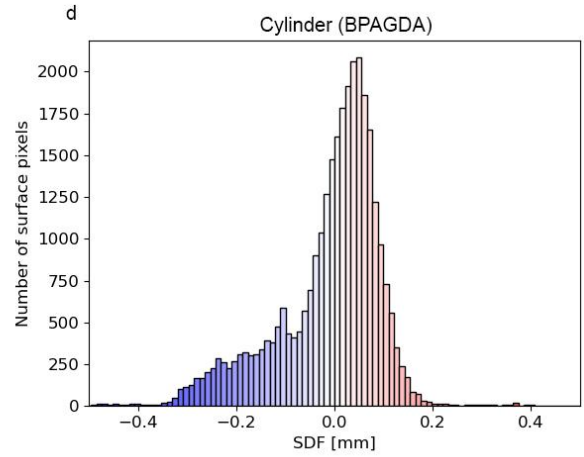
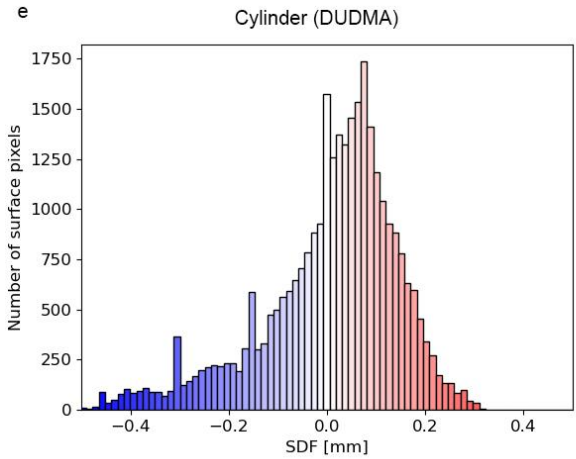
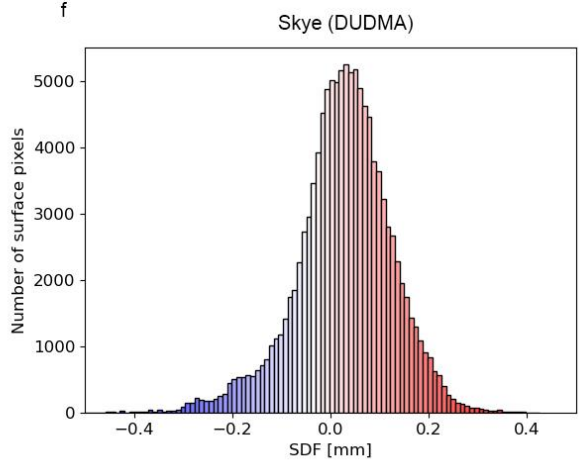

**Fig. S5**. a) – f) SDF (print to reference) histograms for prints in this work.

| | RMSE | RMSE (% of max dimension) | SDF Standard deviation | SDF Standard deviation (% of max dimension) |
|---|---|---|---|---|
| Stanford Bunny (DUDMA) | 0.176mm | 1.45% | 0.162mm | 1.33% |
| Gyroid (BPAGDA) | 0.115mm | 0.81% | 0.103mm | 0.73% |
| Gyroid (BPAGDA, OST to x-ray CT, bottom removed) | 0.172mm | 1.23% | 0.164mm | 1.17% |
| Benchy (BPAGDA) | 0.158mm | 1.28% | 0.110mm | 0.89% |
| 12mm Cylinder (BPAGDA) | 0.112mm | 0.87% | 0.112mm | 0.87% |
| 12mm Cylinder (DUDMA) | 0.141mm | 1.16% | 0.141mm | 1.16% |
| Skye™ (DUDMA, 2x resolution) | 0.102mm | 0.70% | 0.099mm | 0.67% |

**Table S1.** SDF distribution parameters for prints in this work. SDF standard deviations can be compared to those reported in the high fidelity system in [5]. All SDF parameters refer to OST to reference mesh comparison, except otherwise noted for the BPAGDA gyroid print (3rd row, see Fig. S7).

## Section S4: Projection Calculation

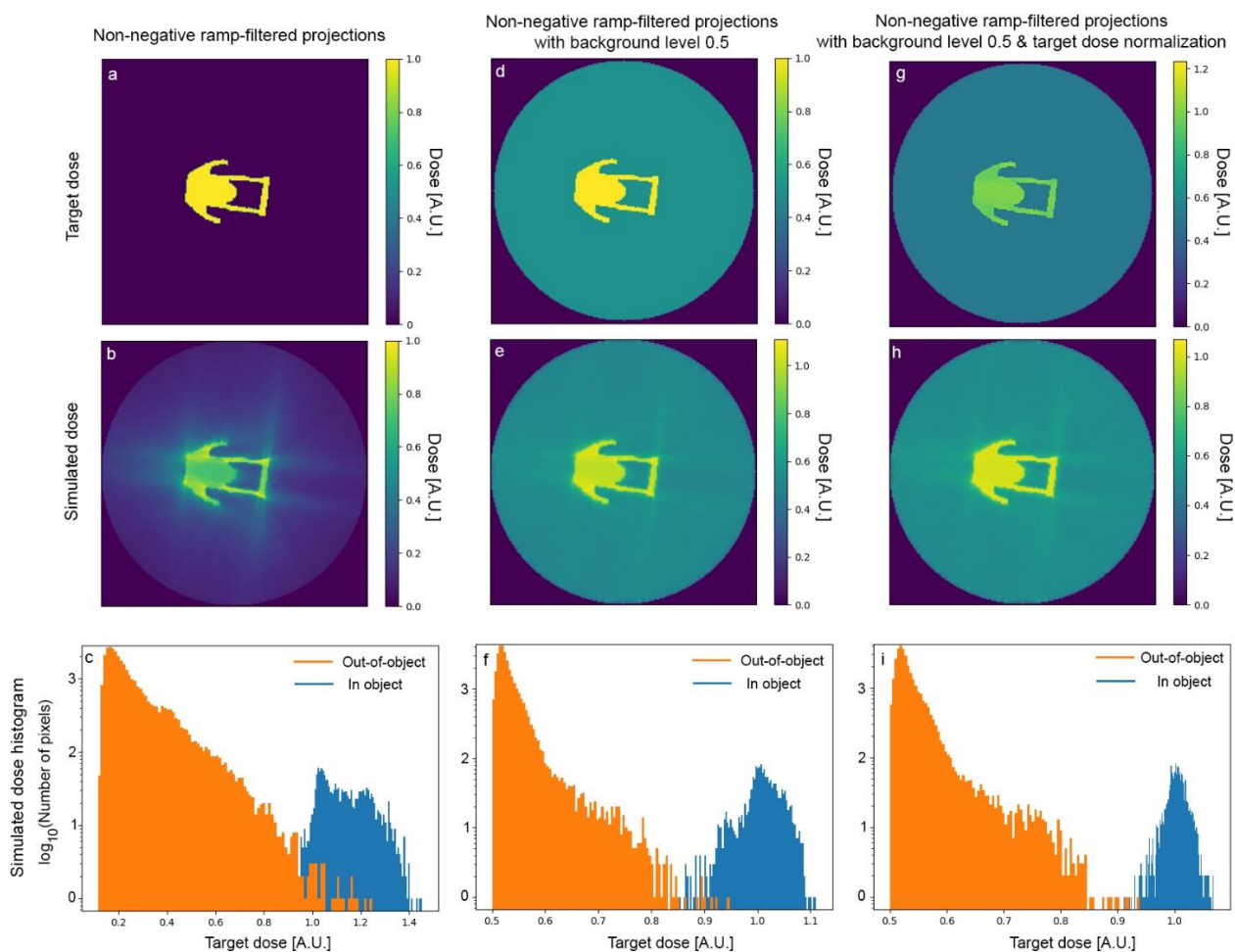

**Fig. S 6**. a)-c) Target dose, simulated dose, and simulated dose histogram for a slice of the Benchy object. d)-f) As in (a)-(c), but with a background level of $B = 0.5$. g)-i) As in (d)-(f) but with one iteration of target dose normalization. For simple non-negative ramp-filtered projections, the simulated dose is uneven and contains artifacts near sharp corners (b). The in-part and out-of-part simulated dose histograms overlap (c), indicating that it is impossible to avoid either overcuring or undercuring some regions. Adding a background level helps to even out the simulated dose (e), and to better separate in-part and out-of-part dose histograms (f). Dose normalization further evens out dose and widens the gap between in-part and-out of-part histograms (i).

## Section S5: X-ray CT

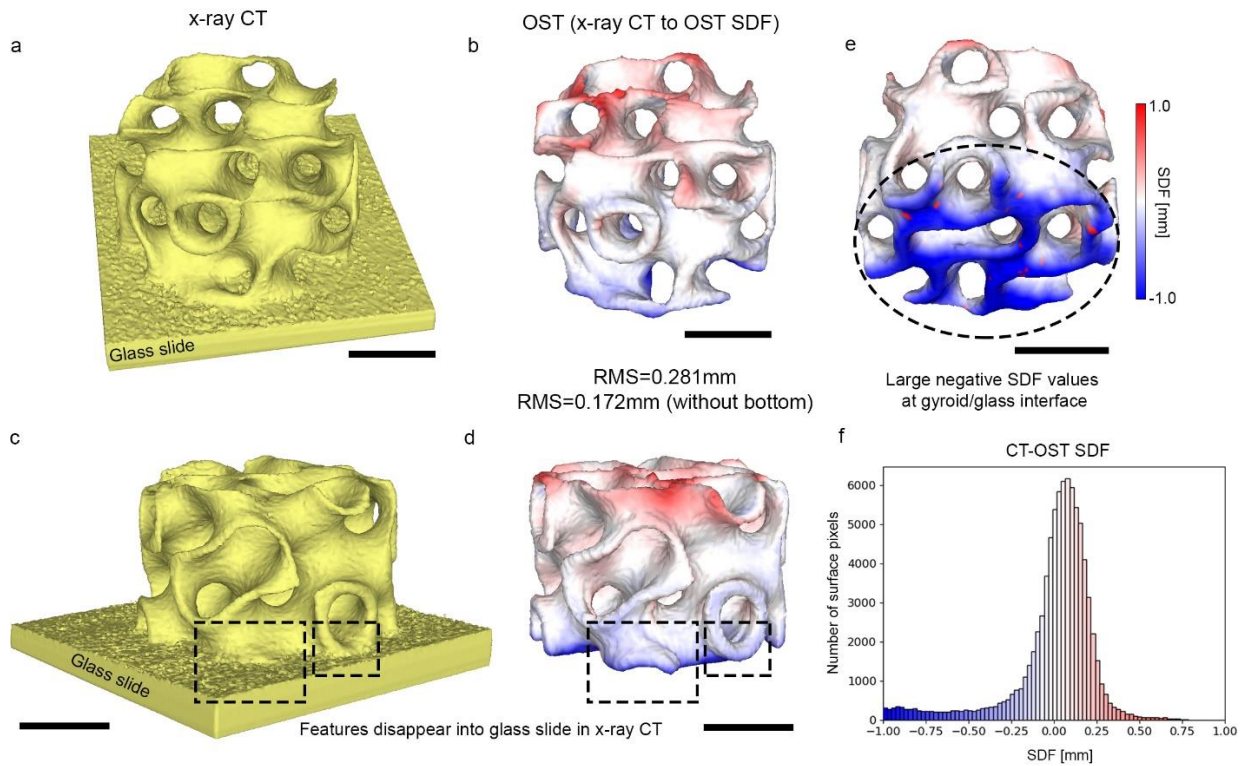

**Fig. S7**. a) Isosurface rendering of x-ray CT scan of the gyroid from Fig. 5e-h. The gyroid is mounted on a glass slide as indicated. b) OST isosurface rendering of the same gyroid. Colormap indicates the SDF between the x-ray CT isosurface in (a) and the OST isosurface. c) As in (a) but for a different viewpoint. Dashed black boxes indicate regions where gyroid features are not visible due to the glass slide. (d) Same as in (b), but for the viewpoint in (c). Note the features highlighted by the dashed black boxes that are not visible in the dashed boxes in the x-ray CT image (c). OST data preserves these features that are otherwise lost at the bottom of the print due to mounting constraints in x-ray CT. e) Bottom view of the SDF, showing large negative values at the bottom of the gyroid. These large negative values occur because matching points in the x-ray CT mesh are not visible due to the glass. The RMS of this OST to x-ray CT is 0.281mm (1.99%) if the bottom of the gyroid is included. If the bottom of the gyroid is ignored, the SDF RMS is 0.172mm (1.21%). The improvement is due to the removal of mesh points that are appear buried in the glass in the x-ray CT. f) Histogram of the CT-to-OST SDF for the gyroid print in this figure. Scale bars are 5mm.

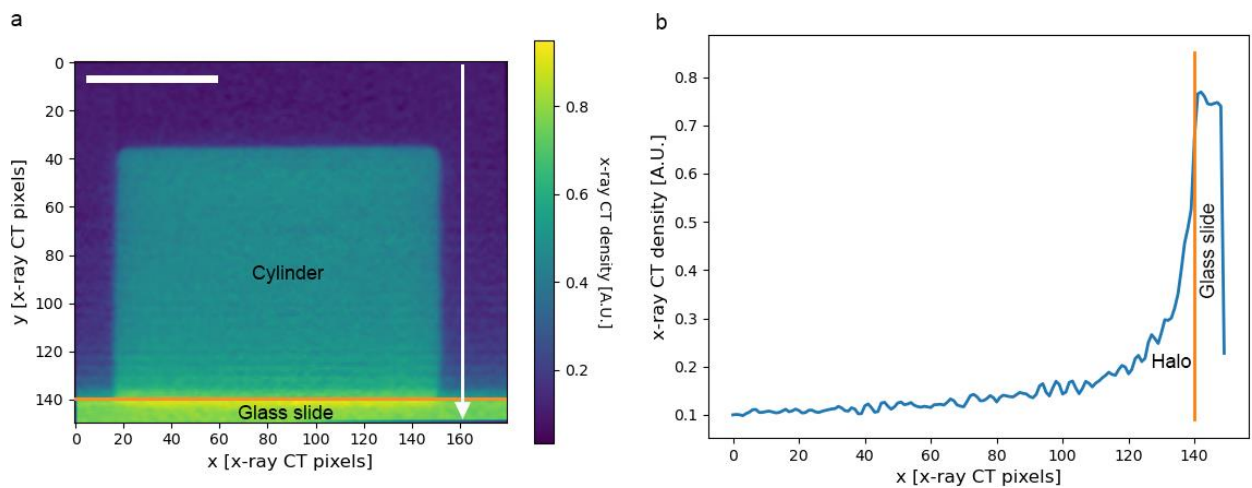

**Fig. S8**. a) X-ray CT image of a slice of the BPAGDA cylinder. The top surface of the glass slide is demarcated by the horizontal orange line. The white line indicates the location and direction of the x-ray CT density plot in (b). Note the increasing x-ray CT density (brightness) along the direction of the white line. We call this a halo resulting from the glass slide used for mounting. (b) X-ray CT density line plot along the white line in (a). The increasing x-ray CT density near the surface of the glass slide is apparent. This halo effect makes isosurfaces from the x-ray CT data prone to errors, in particular overestimating the size of features close to the glass slide. X-ray CT data for samples mounted on a glass slide are not appropriate for metrology due to this artifact. Scale bar is 5mm.

*Section S6: Resolution*

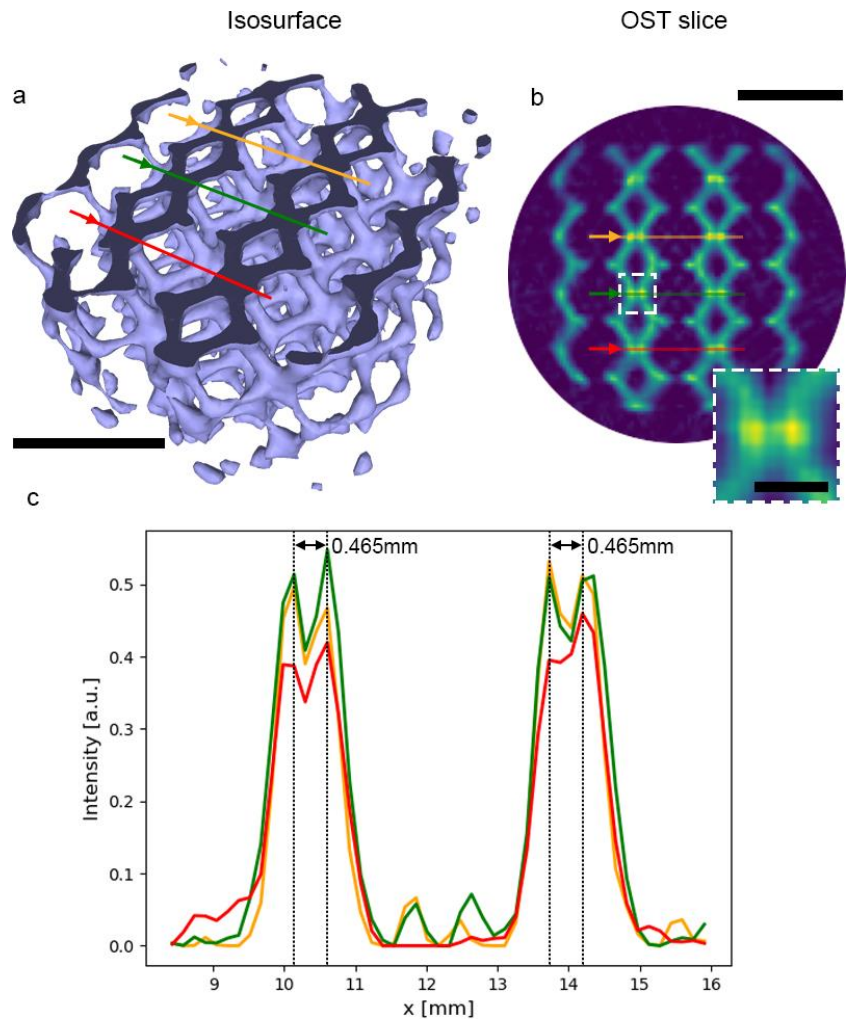

**Fig. S9**. OST resolution characterization for real time imaging experiments. a) OST isosurface rendering of a partially printed lattice structure. Scale bar is 5mm. b) Reconstructed OST slice for the top plane of the cut from the isosurface in (a). Scale bar is 5mm. Inset: Magnified image of boxed white dashed region. Scale bar is 1mm. c)

Intensity line plots along the orange, green and red lines in (a) and (b). Scattering from closely spaced struts is resolved down to three OST pixels (0.465mm).